\documentclass[usenatbib]{mn2e}
\usepackage{mathptmx}
\usepackage{hyperref}
\usepackage{psfig}
\usepackage{epsfig}
\usepackage{pdflscape}
\newif\ifAMStwofonts

\def\sqiglt{\hbox{\rlap{\lower.55ex \hbox {$\sim$}}\kern-.05em \raise.4ex \hbox{$<$}\,}}
\def\sqiggt{\hbox{\rlap{\lower.55ex \hbox {$\sim$}}\kern-.05em \raise.4ex \hbox{$>$}\,}}
\def\til{\ensuremath{\sim\,}}

\newcommand{\tim}[1]{\ensuremath{\times 10^{#1}}}
\def\deg{\ensuremath{^{\circ}}}

\def\cms{\ensuremath{$cm$^{-2}}}

\def\swift{\emph{Swift}}
\def\rosat{\emph{ROSAT}}

\def\t0{\ensuremath{T_{0}}}

\def\arcmin{\ensuremath{^\prime}}
\def\nu{\ensuremath{\nu}}

\title[Swift observations of GW150914]{\swift\ follow-up of the Gravitational Wave source
GW150914}

\author[Evans et al.]{P.A. Evans$^1$\thanks{pae9@leicester.ac.uk}, J.A. Kennea$^2$, S.D. Barthelmy$^3$, A.P. Beardmore$^1$, D. N. Burrows$^2$,
\and S. Campana$^4$, S.B. Cenko$^{3,5}$, N. Gehrels$^3$, P. Giommi$^6$, C. Gronwall$^{2,7}$,
\and F. E. Marshall$^3$, D. Malesani$^8$, C.B. Markwardt$^{3,9}$, B. Mingo$^1$, J. A. Nousek$^2$, 
\and P. T. O'Brien$^1$, J. P. Osborne$^1$, C. Pagani$^1$, K.L. Page$^1$, D.M. Palmer$^{10}$, M. Perri$^{6,11}$,
\and J. L. Racusin$^3$, M.H. Siegel$^2$, B. Sbarufatti$^{2,4}$, G. Tagliaferri$^4$
\\
$^1$ Department of Physics and Astronomy, University of Leicester, Leicester, LE1 7RH, UK \\
$^2$ Department of Astronomy and Astrophysics, Pennsylvania State University, 525 Davey Lab, University Park, PA 16802, USA \\
$^3$ NASA Goddard Space Flight Center, Mail Code 661, Greenbelt, MD 20771, USA \\
$^4$ INAF, Osservatorio Astronomico di Brera, via E. Bianchi 46, 23807 Merate, Italy \\
$^5$ Joint Space-Science Institude, University of Maryland, College Park, MD 20742, USA \\
$^6$ Agenzia Spaziale Italiana (ASI) Science Data Center, I-00133 Roma, Italy; \\
$^7$ Institute of Gravitation and the Cosmos, Institute for Gravitation and the Cosmos, Pennsylvania State University, University Park, PA 16802, USA \\
$^8$ Dark Cosmology Centre, Niels Bohr Institute, University of Copenhagen, Juliane Maries Vej 30, 2100 K\o benhavn \O, Denmark \\
$^9$ Department of Astronomy, University of Maryland, College Park, MD 20742, USA \\
$^{10}$ Los Alamos National Laboratory, B244, Los Alamos, NM, 87545, USA \\
$^{11}$  INAF-Osservatorio Astronomico di Roma, via Frascati 33, I-00040 Monteporzio Catone, Italy \\
}
\date{Accepted -- Received --}
\pagerange{\pageref{firstpage}--\pageref{lastpage}}
\pubyear{2016}

\begin{document}
\maketitle
\label{firstpage}
\begin{abstract} 
The Advanced LIGO observatory recently reported the first direct detection of
gravitational waves (GW) which triggered ALIGO on 2015 September 14. We report on observations taken with the \swift\ satellite two days
after the trigger. No new X-ray, optical, UV or hard X-ray sources were detected 
in our observations, which were focussed on nearby galaxies in the GW error region
and covered 4.7 square degrees (\til2\%\ of the probability in the rapidly-available GW error region; 0.3\%
of the probability from the final GW error region, which was produced several months after the trigger). 
We describe the rapid \swift\ response and automated analysis of the X-ray telescope
and UV/Optical Telescope data, and note the importance
to electromagnetic follow up of early notification of the progenitor details inferred from GW analysis.
\end{abstract}

\begin{keywords}
Gravitional Waves -- Xrays: general -- methods: data analysis
\end{keywords}

\section{Introduction}
\label{sec:intro}
The Advanced LIGO (ALIGO) observatory \citep{Aasi15} recently reported the first ever direct
detection of gravitational waves (GW; \citealt{Abbott16}), ALIGO event GW150914. One of the most likely sources
of GW detectable by ALIGO is the coalesence of a compact binary, i.e.\ one 
containining neutron stars (NS) or stellar-mass black holes (BH). Such events may be accompanied by transient
electromagnetic (EM) radiation such as a short gamma ray burst (`sGRB'; if the binary is viewed close to face-on;
see \citealt{Berger14} for a review)
or a kilonova (see, e.g.\ \citealt{Metzger12,Cowperthwaite15}). Previous searches for coincident EM and GW emission
have produced null results (e.g.\ \citealt{Evans12,Aasi14}).  In a previous work (\citealt{Evans16}; hereafter `Paper I') we discussed
how the \swift\ satellite \citep{GehrelsSwift} could respond to such triggers to search for emission from a short GRB afterglow
with the \swift\ X-ray telescope (XRT; \citealt{BurrowsXRT}).  For GW150914, \swift\ was able
to rapidly respond and was the first EM-facility to report results (\til15 hr after the GW trigger 
was announced, \citealt{EvansLVCCirc1}).

ALIGO uses two approaches to search for GW. The first (`burst') searches for GW signals with
no prior assumptions about the nature of the signal; the second (compact binary coalescence, or CBC) assumes that the signal
comes from the coalescence of a binary comprising neutron stars (NS) and/or black holes (BH), and uses a template
library of expected signals. The `Coherent WaveBurst' (cWB) pipeline, one of the `burst' pipelines,
triggered on 2015 September 14 at 09:50:45 UT, reporting a signal with a false alarm rate of 1.178\tim{-8}\ Hz,
i.e.\ a spurious signal of this significance is expected
once every 2.7 years (this was later revised to less than one every four hundred years; \citealt{LSC_18851} and then
one per 203,000 years, \citealt{Abbott16}).
This event was announced to the EM follow-up partners on 2015 September 16 at 06:39 UT \citep{Singer15a}.
Two skymaps were released originally, one from the cWB pipeline (which uses a likelihood analysis) 
and a refined skymap from the omicron-LALInfereceBursts (`oLIB')
pipeline (which uses a Markov-Chain Monte Carlo [MCMC] approach which is more accurate than 
the cWB approach, but takes longer to perform). For details of these algorithms see \cite{Abbott16d}.
We selected the skymap from the latter, known as `LIB\_skymap', since the LIGO team
reported this as the refined localisation \citep{Singer15a}.

The 90\%\ confidence error region in the `LIB\_skymap' covered 750 square degrees.
In 2016 January  a futher analysis was released by the ALIGO
team. This was produced from the CBC pipeline, since the event was believed to be
a binary coalescence; and  yielded the definitive skymap known as `LALInference'
\citealt{LSC_18858}. This method uses a full MCMC parameter reconstruction. The 90\%\ confidence error region
in this map was reduced to 600 square degrees, \cite{LSC_18858}.

In this letter we report on follow-up observations with the XRT and UV/Optical telescope (UVOT; \citealt{RomingUVOT}),
and we also searched the Burst Alert Telescope (BAT; \citealt{BarthelmyBAT}) data for any sign
of hard X-ray emission at the time of the trigger. A summary of all of the EM follow-up of GW150914 was
given by the LIGO-EM follow up team \citep{Abbott16EM}.

Throughout this paper, errors are quoted at the 90\%\ confidence level unless otherwise stated, and
all fluxes and magnitudes are the observed values (i.e.\ no corrections have been made for 
reddening or absorption by interstellar/intergalactic gas and dust).

\section{\swift\ observations}
\label{sec:response}

\swift\ contains three instruments. The Burst Alert Telescope (BAT, \citealt{BarthelmyBAT}) is a coded mask telescope
with an energy sensitivity of 15--350 keV and field of view \til2 sr. The X-ray Telescope (XRT, \citealt{BurrowsXRT})
covers 0.3--10 keV and has a roughly circular field of view with radius 12.35\arcmin. The UV/Optical telescope
has 7 filters covering the 1270--6240 \AA\ wavelength range, and a square field of view \til17\arcmin\ to a side.
BAT is primarily a trigger instrument not a follow-up instrument and, as discussed in Paper I, the likelihood
of a simultaneous BAT+ALIGO trigger is modest. We used both XRT and UVOT to search for a counterpart to the GW event, noting
(as per Paper I) that the expected rate of unrelated X-ray transients serendipitously detected is lower than in the optical
bands, therefore our strategy is optimised for the XRT.

While Paper I advocated a large-scale rapid tiling with \swift\ in reponse to a GW trigger,
such an operating mode had not been commissioned when ALIGO triggered on GW150914\footnote{Indeed, the trigger
actually occured at the end of an engineering run, before the official start of the O1 observing run.}, therefore we 
were obliged to observe a smaller number of fields. Following Paper I we convolved the ALIGO
sky localisation map (we used the `LIB\_skymap' which was the best map available at time of our observations; \citealt{Abbott16})
with the Gravitational Wave Galaxy Catalogue (GWGC; \citealt{White11}). 
We added a 100 kpc halo to each galaxy in the GWGC more than 5 Mpc away\footnote{For
galaxies closer than this 5 Mpc, the angular
projection of a 100 kpc halo covers an unreasonably large fraction of
the sky, but the fraction of binary NS mergers which occur within this distance is negligible: 0.013\%\ assuming
they are homogeneously distributed in space and detectable to 100 Mpc by ALIGO.}, to 
reflect possible impact of natal NS kicks.  However, unlike our previous work (where pixels in this map had values of 1 or 0) we weighted this map
by galaxy luminosity. Each GWGC galaxy was assigned a probability $P=L/L_{\rm tot}$, where $L$ is the 
B-band luminosity of the galaxy reported in the GWGC and $L_{\rm tot}$ is the total luminosity of all galaxies
in the catalogue. This probability was then evenly distributed between all {\sc healpix}\footnote
{Hierarchical Equal Area isoLatitude Pixelization \protect\citep{Gorski05}, the file format in which ALIGO 
error regions are disseminated.} pixels 
corresponding to the galaxy and its halo, i.e. we assumed that the probability of a binary neutron star merger
(which gives rise to the GRB and GW emission) is spatially uniform throughout the host galaxy and its halo.

\begin{figure}
\begin{center}
\includegraphics[width=8.1cm,angle=0]{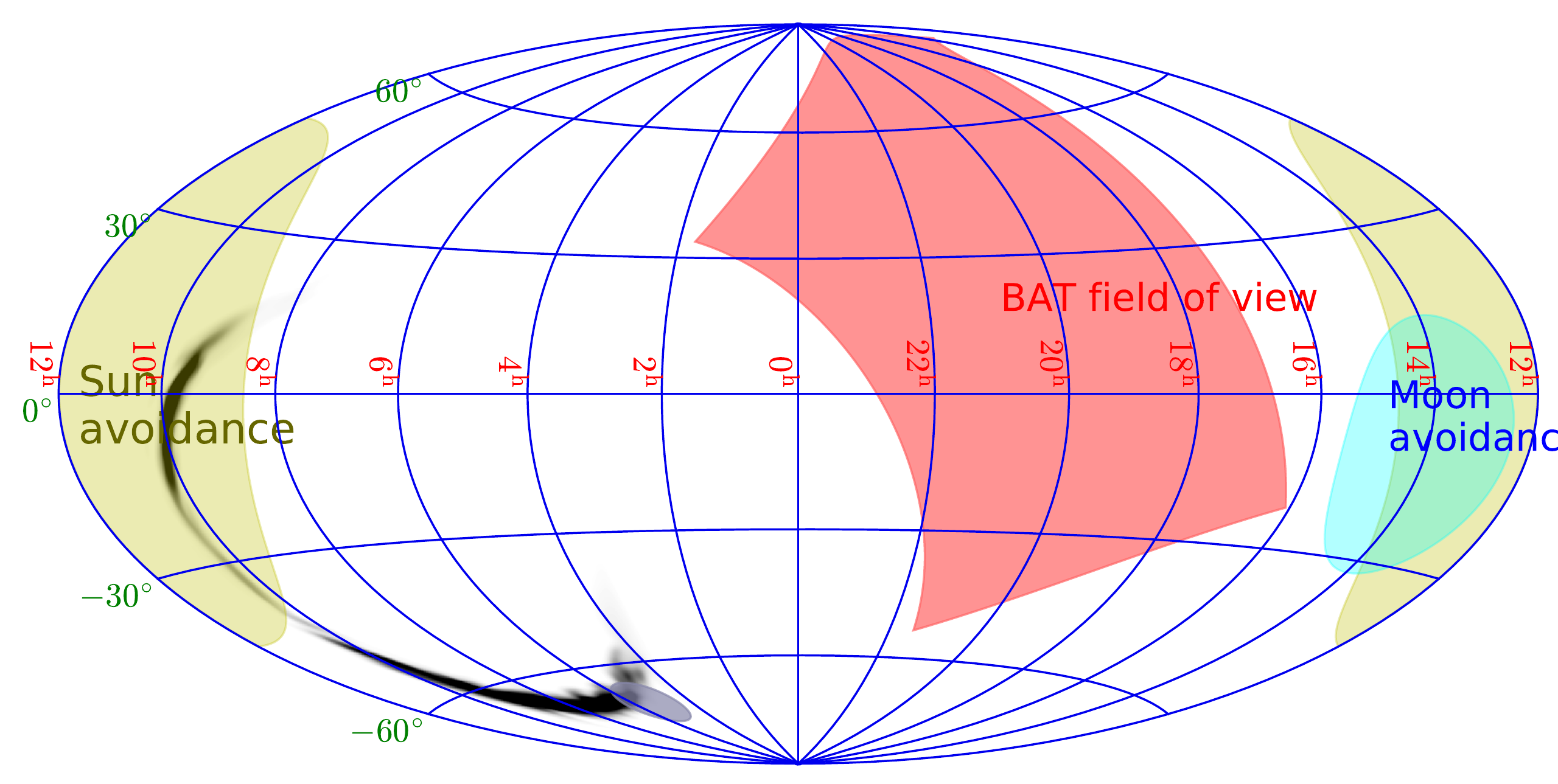}
\includegraphics[width=8.1cm,angle=0]{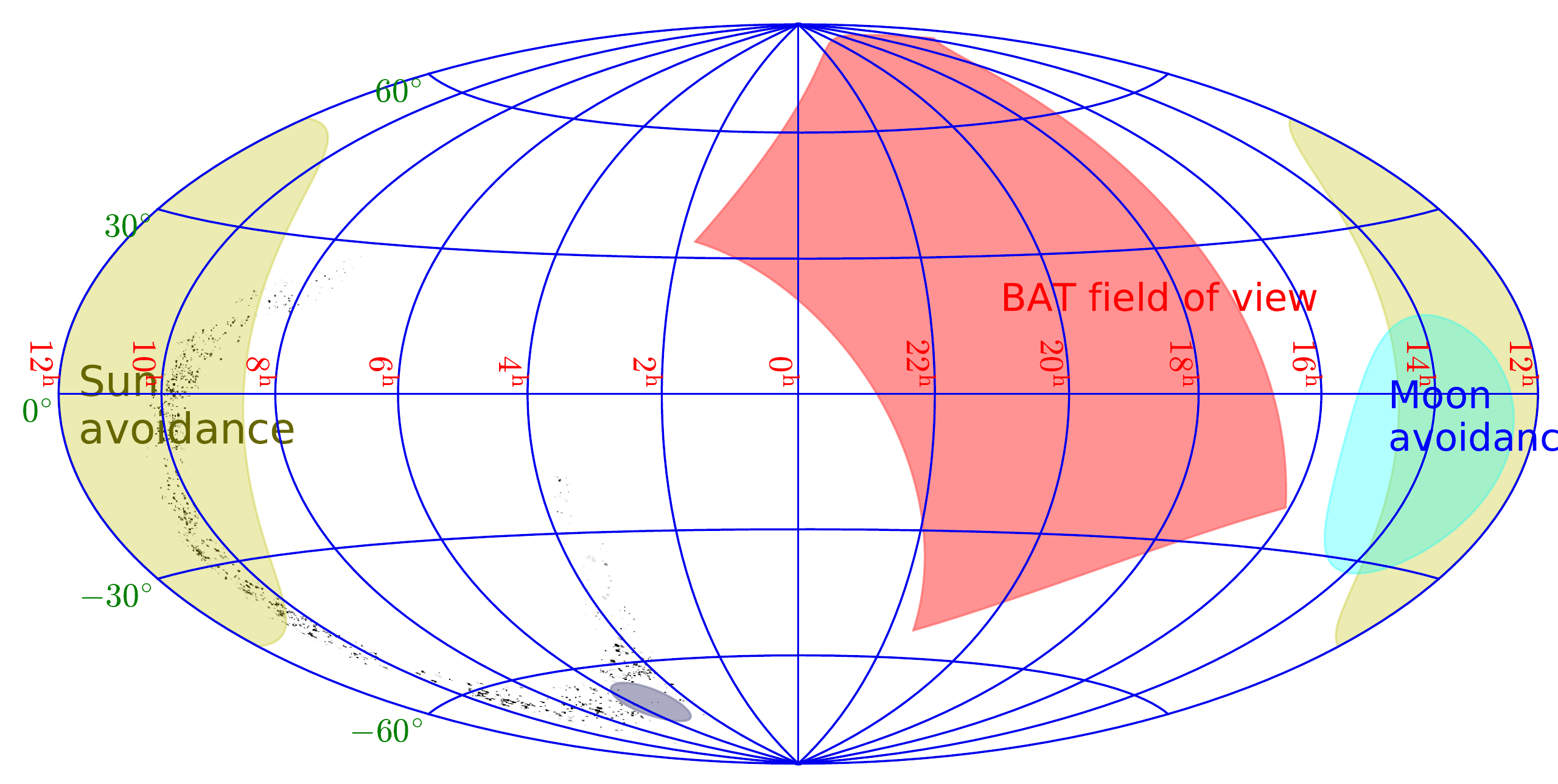}
\includegraphics[width=8.1cm,angle=0]{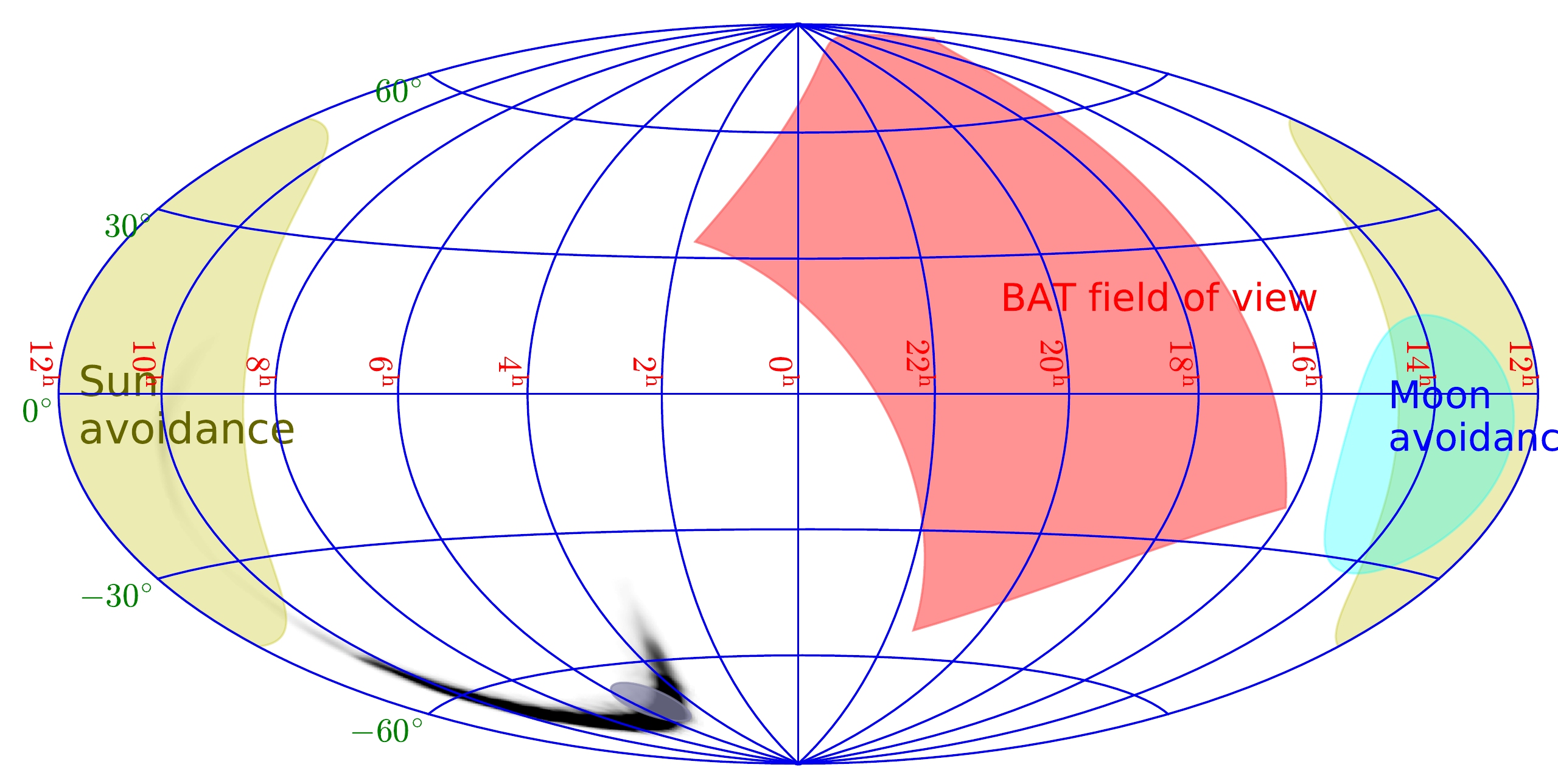}
\end{center}
\caption{The top two panels show the `LIB\_skymap' GW localisation map produced by the LVC team on 2015 September 15, in the
original form (top) and convolved with our
luminosity-weighted GWGC map (middle). The bottom panel shows the revised `LALInference' skymap released on 2016 January 13.
Coordinates are equatorial, J2000. The yellow and cyan circles show the regions of the sky which \swift\ could not
observe due to the presence of the Sun and Moon respectively, calculated at the time of the first \swift\ observations.
The small, lilac ellipse marks the LMC. The large purple region approximates the BAT field of view at the time of the GW trigger.}
\label{fig:150914_skymap}
\end{figure}

The LIB\_skymap from the ALIGO team and the version convolved 
with GWGC are shown in Fig.~\ref{fig:150914_skymap}, along with the final skymap, released in 2016 January.

Unfortunately
a large portion of the error region was within \swift's Sun observing constraint where \swift's narrow field instruments cannot be pointed ($<47\deg$ from the Sun).
\cite{Gehrels16} noted that, since most of the galaxy luminosity comes from a small fraction of galaxies, 
one can opt to only observe the brightest galaxies within the ALIGO error region; \cite{Kasliwal15}
reported a list of such galaxies. Unfortunately, three of their top ten were within the Sun constraint region.
It later transpired that this was less of an issue than believed at the time of the GW event, as
the probability contained within the Sun constraint region was significantly reduced in the final
skymap, as the bottom panel of Fig.~\ref{fig:150914_skymap} shows.

\begin{figure}
\begin{center}
\includegraphics[width=8.1cm,angle=0]{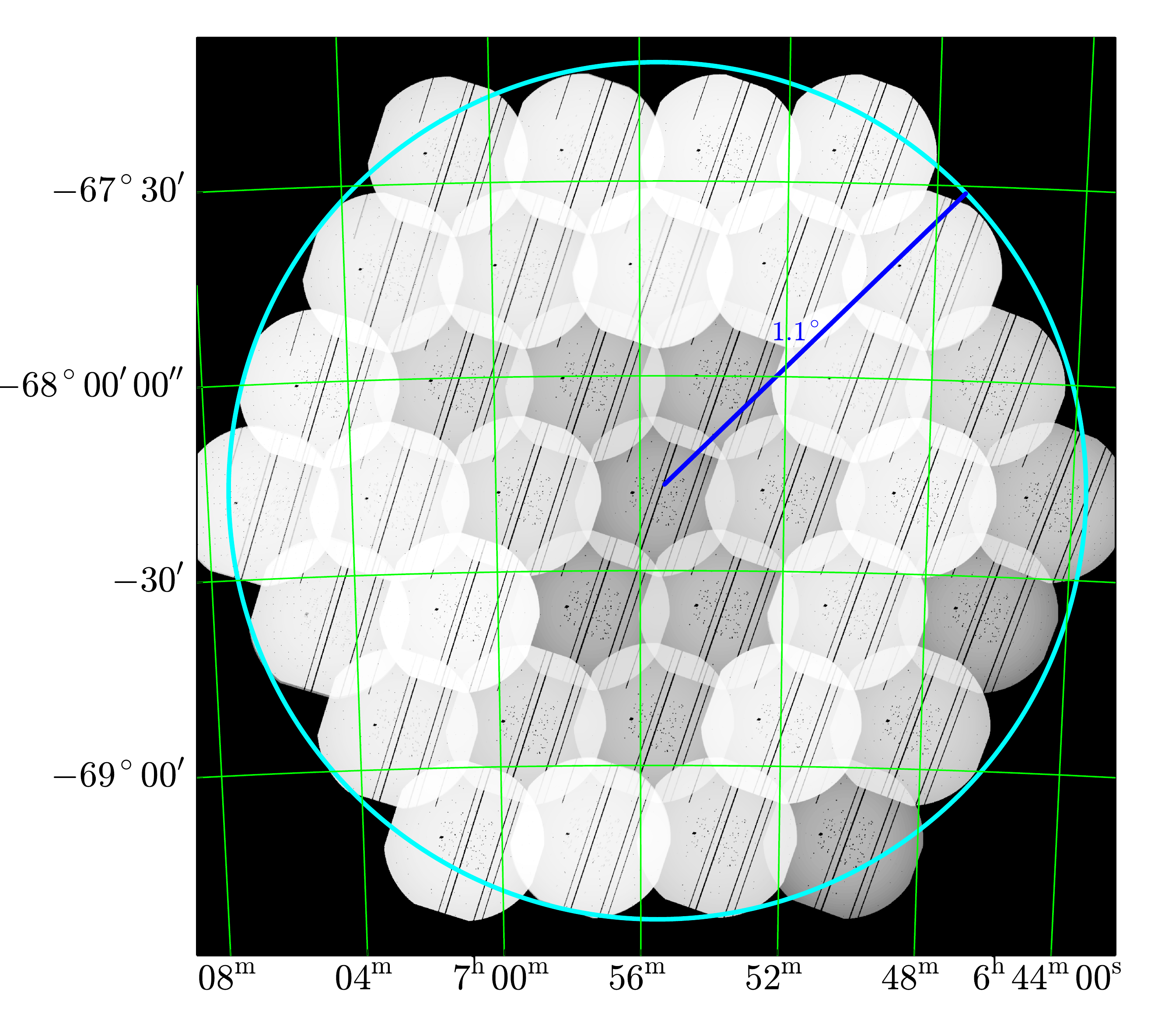}
\end{center}
\caption{The XRT exposure map of the 37-point tiled observations of the LMC performed with \swift,
demonstrating the structure of the pattern. The black lines are the vetoed columns on the CCD.
The cyan circle has radius of 1.1\deg\ and is shown for reference. Axis are RA and Dec, J2000.}
\label{fig:LMC_map}
\end{figure}

Rather than select where to point the XRT based on individual galaxies, we divided the observable sky into XRT fields
of view (since one field may contain multiple galaxies) -- circles of radius 12.35\arcmin\ -- 
and ranked them in decreasing order of probability derived from our GWGC-convolved skymap, and then
observed the top five fields from this list, which contained eight GWGC galaxies.
Noting that the GW error region also intersected the Large Magellanic Cloud (LMC),
we also performed a 37-point tiled observation focussed on the LMC (Fig.~\ref{fig:LMC_map}). This tiling formed part
of the process of commissioning the ability to observe as Paper I advocated, and as it was the first
such test, we were limited to short exposures. A full list of the \swift\ observations is given in Table~\ref{tab:obs_150914}.

\begin{table}
\begin{center}
\caption{\swift\ observations of the error region of GW150914}
\label{tab:obs_150914}
\begin{tabular}{ccc}
\hline
Pointing direction & Start time$^a$ & Exposure \\
(J2000) & (UTC) & (s) \\
\hline
$09^h 13^m 29.65^s,-60\deg 43^\prime 37.4^{\prime\prime}$  &  Sep 16 at 15:19:27  &   777  \\
$08^h 16^m 30.77^s,-67\deg 38^\prime 06.7^{\prime\prime}$  &  Sep 16 at 16:54:41  &   987  \\
$07^h 28^m 42.38^s,-66\deg 59^\prime 43.1^{\prime\prime}$  &  Sep 16 at 18:28:32  &   970  \\
$08^h 03^m 23.72^s,-67\deg 37^\prime 17.2^{\prime\prime}$  &  Sep 16 at 20:05:37  &   970  \\
$08^h 57^m 17.34^s,-65\deg 26^\prime 34.1^{\prime\prime}$  &  Sep 16 at 21:42:15  &   985  \\
\hline
LMC Observations \\

$06^h 55^m 30.59^s,-68\deg 18^\prime 44.3^{\prime\prime}$  &  Sep 17 at 18:26:54  &   20  \\
$06^h 59^m 13.43^s,-68\deg 18^\prime 29.7^{\prime\prime}$  &  Sep 17 at 18:28:03  &   42  \\
$06^h 57^m 21.25^s,-68\deg 36^\prime 12.8^{\prime\prime}$  &  Sep 17 at 18:29:12  &   20  \\
$06^h 53^m 42.84^s,-68\deg 36^\prime 04.4^{\prime\prime}$  &  Sep 17 at 18:30:21  &   22  \\
$06^h 51^m 53.97^s,-68\deg 18^\prime 16.7^{\prime\prime}$  &  Sep 17 at 18:31:29  &   32  \\
$06^h 53^m 45.48^s,-68\deg 00^\prime 43.4^{\prime\prime}$  &  Sep 17 at 18:32:38  &   22  \\
$06^h 57^m 25.10^s,-68\deg 01^\prime 02.6^{\prime\prime}$  &  Sep 17 at 18:33:46  &   25  \\
$07^h 01^m 1.84^s,-68\deg 01^\prime 05.6^{\prime\prime}$  &  Sep 17 at 18:34:54  &   35  \\
$07^h 02^m 52.89^s,-68\deg 18^\prime 56.6^{\prime\prime}$  &  Sep 17 at 18:36:02  &   72  \\
$07^h 01^m 0.50^s,-68\deg 36^\prime 16.1^{\prime\prime}$  &  Sep 17 at 18:37:09  &   82  \\
$06^h 59^m 11.14^s,-68\deg 53^\prime 42.6^{\prime\prime}$  &  Sep 17 at 18:38:17  &   37  \\
$06^h 55^m 32.45^s,-68\deg 53^\prime 32.4^{\prime\prime}$  &  Sep 17 at 18:39:25  &   25  \\
$06^h 51^m 54.75^s,-68\deg 53^\prime 32.0^{\prime\prime}$  &  Sep 17 at 18:40:33  &   65  \\
$06^h 50^m 5.28^s,-68\deg 35^\prime 51.8^{\prime\prime}$  &  Sep 17 at 18:41:40  &   52  \\
$06^h 48^m 15.62^s,-68\deg 18^\prime 20.6^{\prime\prime}$  &  Sep 17 at 18:42:47  &   65  \\
$06^h 50^m 6.94^s,-68\deg 00^\prime 54.0^{\prime\prime}$  &  Sep 17 at 18:43:53  &   60  \\
$06^h 51^m 56.98^s,-67\deg 43^\prime 22.9^{\prime\prime}$  &  Sep 17 at 18:44:59  &   67  \\
$06^h 55^m 34.08^s,-67\deg 43^\prime 36.1^{\prime\prime}$  &  Sep 17 at 18:46:04  &   72  \\
$06^h 59^m 13.52^s,-67\deg 43^\prime 33.4^{\prime\prime}$  &  Sep 17 at 18:47:10  &   55  \\
$07^h 02^m 51.97^s,-67\deg 43^\prime 41.4^{\prime\prime}$  &  Sep 17 at 18:48:15  &   62  \\
$07^h 04^m 42.41^s,-68\deg 01^\prime 15.1^{\prime\prime}$  &  Sep 17 at 18:49:21  &   75  \\
$07^h 06^m 30.83^s,-68\deg 18^\prime 50.4^{\prime\prime}$  &  Sep 17 at 18:50:27  &   70  \\
$07^h 04^m 41.09^s,-68\deg 36^\prime 37.2^{\prime\prime}$  &  Sep 17 at 18:51:32  &   60  \\
$07^h 02^m 50.35^s,-68\deg 53^\prime 43.9^{\prime\prime}$  &  Sep 17 at 18:52:38  &   60  \\
$07^h 01^m 1.00^s,-69\deg 11^\prime 19.8^{\prime\prime}$  &  Sep 17 at 18:53:43  &   62  \\
$06^h 57^m 21.83^s,-69\deg 11^\prime 05.0^{\prime\prime}$  &  Sep 17 at 18:54:49  &   67  \\
$06^h 53^m 43.60^s,-69\deg 11^\prime 06.9^{\prime\prime}$  &  Sep 17 at 18:55:55  &   42  \\
$06^h 50^m 4.65^s,-69\deg 11^\prime 01.6^{\prime\prime}$  &  Sep 17 at 20:02:45  &   20  \\
$06^h 48^m 14.61^s,-68\deg 53^\prime 22.8^{\prime\prime}$  &  Sep 17 at 20:03:54  &   32  \\
$06^h 46^m 25.66^s,-68\deg 35^\prime 44.9^{\prime\prime}$  &  Sep 17 at 20:05:02  &   20  \\
$06^h 44^m 35.32^s,-68\deg 18^\prime 21.1^{\prime\prime}$  &  Sep 17 at 20:06:11  &   25  \\
$06^h 46^m 27.88^s,-68\deg 00^\prime 48.6^{\prime\prime}$  &  Sep 17 at 20:07:19  &   35  \\
$06^h 48^m 17.47^s,-67\deg 43^\prime 23.8^{\prime\prime}$  &  Sep 17 at 20:08:27  &   60  \\
$06^h 50^m 7.30^s,-67\deg 25^\prime 50.9^{\prime\prime}$  &  Sep 17 at 20:09:34  &   70  \\
$06^h 53^m 44.83^s,-67\deg 26^\prime 05.6^{\prime\prime}$  &  Sep 17 at 20:10:41  &   77  \\
$06^h 57^m 24.51^s,-67\deg 26^\prime 04.1^{\prime\prime}$  &  Sep 17 at 20:11:48  &   67  \\
$07^h 01^m 2.66^s,-67\deg 26^\prime 08.1^{\prime\prime}$  &  Sep 17 at 20:12:54  &   57  \\
\hline
\end{tabular}
\end{center}
\medskip
$^a$ All observations were in 2015.
\end{table} 

\section{Data Analysis}
\label{sec:analysis}
The XRT data analysis was largely automated, using custom software produced by (and running at) the UK Swift Science Data Centre at the 
University of Leicester. This software makes extensive use of the \swift\ software\footnote{V4.3, part of {\sc heasoft} 6.15.1.}
with the latest CALDB files\footnote{Released on 2015 July 21.}.

Our baseline source detection system was that developed by \cite{Evans14} for the \swift-XRT Point Source (1SXPS) catalogue.
This is an iterative system, that uses a sliding-cell detection method with a background map which
is recreated on each pass to account for sources already discovered. The majority of datasets in the 1SXPS catalogue corresponded to a single XRT field of view, and even where
multiple fields of view overlapped, images were limited to 1000$\times$1000 pixels ($39.3\arcmin\times39.3\arcmin$)
in size. For GW150914 we observed a large contiguous region (part of the LMC; Fig.~\ref{fig:LMC_map}) -- this region is so large
that the background mapping developed for 1SXPS is not properly calibrated, and the coordinates
in the tangent-plane projection become inaccurate. We therefore broke the data into `analysis blocks'. Each block was no more than 0.55\deg\ in radius
(equivalent to a 7-point tile), and every XRT field of view had to be in at least one block. Any redundant blocks
(i.e. where every XRT field in the block was also in another block) were removed. Since this meant that some 
areas of sky were in multiple blocks, we checked for duplicate detections of the same source (based on spatial coincidence) from multiple
blocks and merged any that occured.

In 1SXPS the minimum exposure time permitted was 100 s, however Table~\ref{tab:obs_150914} shows that many of the 
observations (those of the LMC) were shorter than this. This is not likely to be a problem regarding spurious source detections; \cite{Evans14} found that
at short exposure times there are very few spurious sources due to the lack of background events. However,
this lack of background may mean that we can reduce the signal-to-noise ratio (S/N) threshold for
sources to be accepted in short exposure. We simulated 50-s exposure images (both single fields,
and 7-point tiles, to represent the extreme sizes of the analysis blocks) in a manner analagous to \cite{Evans14}
and found that for an S/N threshold of 1.3, the rate of spurious detections was $<3/1000$, equivalent to the
`Good' flag in 1SXPS; this was therefore used for the short ($<100$ s) observations.

As discussed in Paper I, the discovery of an X-ray source alone does not identify it as the counterpart to the GW trigger.
We therefore gave each detected source a `rank' indicating how likely it is to be related to the GW event, from 1 (very likely)
to 4 (very unlikely). This involved comparing our source detections with the \emph{ROSAT} All Sky Survey (RASS; \citealt{Voges99}).
To do this we assumed a typical AGN spectrum, a power-law with hydrogen
column density $N_H=3\tim{20}$ \cms\ and a photon index of $\Gamma=1.7$.
These ranks were defined as follows:

\noindent\emph{Rank 1: Good GW counterpart candidate.} Sources which lie within 200 kpc of a GWGC
galaxy, and are either uncatalogued and brighter than the 3-$\sigma$ catalogue limit, or catalogued
but brighter than their catalogued flux. In both cases, `brighter than' means that the measured and historical
values (or upper limits) disagree at the 5-$\sigma$ level. For uncatalogued sources, the comparison is to
the RASS, or to 1SXPS or the \emph{XMM-Newton} catalogues, if an upper limit from those catalogues is available and deeper than the RASS limit.

\noindent\emph{Rank 2: Possible counterpart.} The criteria for this are similar to those above, except that `brighter' is determined
at the 3-$\sigma$ level, and there is no requirement for the source to be near a known galaxy.

\noindent\emph{Rank 3: Undistinguished source.} Sources which are uncatalogued, but are fainter
than existing catalogue limits, or consistent with those limits at the 3-$\sigma$ level. i.e.\ sources which
cannot be distinguished from field sources.

\noindent\emph{Rank 4: Not a counterpart.} Sources which are catalogued, and which have fluxes consistent with (at the 3-$\sigma$ level)
or fainter than their catalogued values.

The relatively conservative flux requirements of rank 1 arise because of biases which cause us to overestimate
the ratio between the observed flux and historical flux or limit. The Eddington bias \citep{Eddington40} results in
the fluxes of sources close to the detection limit being overestimated; this was discussed and quantified
for \swift-XRT by \cite{Evans14} (section 6.2.1 and figs.~9--10). Also, \rosat\ had a much softer response
than \swift\ (0.1--2.4 keV compared to 0.3--10 keV); for sources with harder spectra than in our
assumed model (especially those more heavily absorbed) this means that \rosat\ was less sensitive, i.e.\
our calculated XRT/\rosat\ ratio will be too low.

As well as the checks performed to automatically rank each XRT source, the 2MASS catalogue \citep{Skrutskie06}
and SIMBAD database \citep{Wenger00} were automatically searched, and any sources within the 3-$\sigma$ XRT error region
were identified. This information was not used to determine the source rank, but to inform human decisions
as to the nature of the source. It is important to note that this spatial correlation does not
necessarily mean that the XRT source and the 2MASS/SIMBAD object are the same thing: \cite{Evans14} showed that
\til11\%\ of XRT sources with SIMBAD matches, and \til64\%\ of those with 2MASS matches are not related but chance
alignments. 

An automated pipeline was built to search for candidate counterparts in 
the UVOT observations using standard {\sc heasoft} analysis tools. In 
the pipeline the tool {\sc uvotdetect} was used to search for sources in 
the sky image files. For each observation searches were made using the 
longest exposure and the sum of all images if the summed exposure was 
significantly longer than the longest exposure. Candidate sources whose 
images were not star-like or were too close to other sources were 
rejected. Sources without counterparts in the USNO-B1.0 catalogue \citep{Monet03}
or Hubble Guide Star Catalog \citep{Lasker08} were considered possible candidates. 
The UVOT image near each of these possible candidates was then visually 
compared with the corresponding region in the Digitized Sky Survey. This visual 
comparison was used to reject candidates due to readout streaks or ghost 
images of bright sources.

The UVOT images near Rank 1 or Rank 2 XRT sources were also
examined and compared with the DSS. UVOT source magnitudes
or upper limits were determined using the tool {\sc uvotsource}.

\section{Results}
\label{sec:results}

\begin{table*}
\begin{center}
\caption{Sources detected by \swift-XRT in follow-up of GW150914, with $u$-band magnitudes from UVOT.}
\label{tab:res_150914}
\begin{tabular}{cccccc}
\hline
RA & Dec & Error & Flux & $u$ Magnitude & Catalogued name \\
(J2000) & (J2000) & 90\%\ conf. & 0.3--10 keV, erg \cms\ s$^{-1}$ & AB mag \\
\hline
09h 14m 06.54s & -60\deg32$^\prime$ 07.7$^{\prime\prime}$ & 4.8$^{\prime\prime}$ & $(1.9\pm0.5)\tim{-12}$ & N/A & XMMSL1 J091406.5-603212 \\
09h 13m 30.24s & -60\deg47$^\prime$ 18.1$^{\prime\prime}$ & 6.1$^{\prime\prime}$ & $(5.3\pm2.0)\tim{-13}$& 15.44$\pm$0.02$^a$ & ESO 126-2 = 1RXS J091330.1-604707 \\
08h 17m 60.62s & -67\deg44$^\prime$ 03.9$^{\prime\prime}$ & 4.7$^{\prime\prime}$ & $(8.9\pm2.4)\tim{-13}$ & 17.53$\pm$0.05 & 1RXS J081731.6-674414  \\
\hline
\end{tabular}
\end{center}
\medskip
$^a$ Magnitude of the core. The galaxy as a whole (removing foreground stars) has a $u$ magnitude of 14.15$\pm$0.02.
\end{table*}

Three X-ray objects were found in the initial observations (the five most probable XRT fields, Section~\ref{sec:response}) and announced
by \cite{EvansLVCCirc1}. These were all known X-ray emitters showing no sign of outburst and assigned a rank of 4, see Table~\ref{tab:res_150914} for details.
XMMSL1 J091406.5-603212 was automatically flagged as being potentially
spurious due to optical loading as it is spatially coincident with HD 79905 which SIMBAD reports as a B9.5 star \citep{Houk75}
with a V magnitude of 7.436 \citep{Kiraga12}, above the threshold where optical loading is likely
to affect the X-ray measurements\footnote{http://www.swift.ac.uk/analysis/xrt/optical\_tool.php}. The measured
count-rate ($0.045\pm0.011$ ct s$^{-1}$) is however slightly lower than that in the RASS ($0.10\pm0.01$ ct s$^{-1}$ when converted
to an XRT-equivalent rate using the AGN spectrum introduce above), suggesting that the optical loading has not resulted in a spurious X-ray detection.
No other SIMBAD objects match the position of this source.
ESO126-2 is listed as an AGN by SIMBAD, whereas 1RXS J081731.6-674414 is simply listed as an X-ray source.

The 3-$\sigma$ upper limit on any other X-ray point source in the initial five fields is 1.5\tim{-2} ct s$^{-1}$, which correponds
to a flux of 6.5\tim{-13} erg \cms\ s$^{-1}$, assuming the AGN spectrum defined above.
For the LMC observations the typical upper limit was 0.16 ct s$^{-1}$, or 6.9\tim{-12} erg \cms\ s$^{-1}$, corresponding
to a luminosity of 2.0\tim{36} erg s$^{-1}$.

UVOT observations were all carried out in the $u$ filter, 2 of the sources were detected and one lay outside the UVOT field of view.
Details are in Table~\ref{tab:res_150914}.
No transient sources were detected by UVOT down to an AB magnitude of \til19.8 for the initial 5 galaxies, and ~18.8 for the LMC.

Both the rapidly available ALIGO sky localisation and the later, revised version
had no probability within  \swift-BAT's nominal field of view,
therefore the lack of a simultaneous trigger from BAT is not informative. 
However, BAT can detect GRBs from outside of the field of view,
by means of gamma-rays that leak through the sidewall shielding.  A search for any corresponding
rate increases from a correlated GRB within $\pm$100 s of the GW signal found no peaks above
the 3-$\sigma$ value of 200 ct s$^{-1}$ above background in the nominal 50--300 keV energy range at a 1-s timescale.
To convert this to a flux limit requires precise knowledge of the direction in which the event occured which
we do not have for GW 150914.

\emph{Fermi}-GBM reported a possible low significance gamma-ray event temporally coincident with the ALIGO trigger
\citep{Blackburn15,Connaughton16}, although this was not detected by INTEGRAL \citep{Ferrigno15} and
no signal was seen in BAT either. The best position deduced by \cite{Connaughton16} was below the Earth 
limb from the perspective of BAT, so the lack of signal is perhaps not surprising; however the GBM localisation
is very poor, covering thousands of square degrees. As just noted, this prevented us from creating an accurate
flux limit for BAT; however, considering the range of possible angles, an approximate 5-$\sigma$ upper
limit over the 14--195 keV band is \til2.4$\tim{-6}$ erg \cms\ s$^{-1}$. \cite{Connaughton16} fit the spectrum
of the GBM event as a power-law with a photon index of 1.4; the fluence from this spectrum
was $2.4^{+1.7}_{-1.0}\tim{-7}$ erg \cms. Since the duration of the pulse was 1 s, the flux has the same
numeric value. This spectrum gives a flux of $7.63\tim{-8}$ erg \cms\ s$^{-1}$; below the upper limit 
derived from BAT. Therefore even if the GBM detection was a real astrophysical event,
it was likely too faint for BAT to have detected, given that the source was outside the coded field of view.

\section{Discussion and conclusions}
\label{sec:disc}

The XRT observations covered 4.7 square degrees, and contained 2\%\ of the probability from the original `LIB\_skymap'
ALIGO error region (8\%\ if this is convolved with the GWGC), and were obtained from 53.5 to 82.3 hr after the GW trigger.
However, \cite{Abbott16} reported
that the most likely source of the GW event is a binary black-hole trigger at 
500 Mpc. Since the GWGC only extends to 100 Mpc and the coalescence of two stellar mass black holes
is not expected to produce EM radiation, our lack of detection is not surprising. Additionally, the recently-released
revised skymap `LALInference' contains much less probability at the location of the XRT fields, with those
field containing only 0.3\%\ of the GW probability (this figure does not change with galaxy convolution).

The possible detection of an sGRB coincident with the ALIGO trigger reported by \emph{Fermi}-GBM is
intriguing, but unfortunately we are not able to place any meaningful constraints on its brightness with the
BAT. None of the GW probability was within the BAT field of view, and the flux limits we can derive for emission
received through the sidewalls of the instrument are above the level expected from the GBM data.

Although the \swift\ observations did not yield the detection of an EM counterpart to the GW trigger, we have demonstrated that \swift\ 
is able to respond very rapidly to GW triggers with \swift: the 3 X-ray sources we detected were 
reported to the GW-EM community within 15 hours of the trigger being announced. In the event
of a nearby binary neutron-star merger triggering ALIGO, such rapid response, analysis and dissemination will be vital.
It is also evident that the decisions made regarding where to observe with \swift\ are best informed if details
such as estimated distances and masses are available rapidly from the GW teams, as noted by the GW-EM summary paper
\citep{Abbott16EM}, and it is expected that the latencies in deriving these parameters will be reduced in the future. We have also
commissioned new observing modes with \swift\ which will allow us to perform much more extensive follow-up
observations of future GW triggers.

\section*{Acknowledgements}
This work made use of data supplied by the UK Swift Science Data Centre 
at the University of Leicester, and used the ALICE High Performance 
Computing Facility at the University of Leicester.  This research has made use of the  XRT Data 
Analysis Software (XRTDAS) developed under the responsibility of the ASI 
Science Data Center (ASDC), Italy. PAE, APB, BM, KLP and JPO acknowledge UK Space Agency support. 
SC and GT acknowledge Italian Space Agency support.
This publication makes use of data products from the Two Micron All Sky Survey, which is 
a joint project of the University of Massachusetts and the Infrared 
Processing and Analysis Center/California Institute of Technology, 
funded by the National Aeronautics and Space Administration and the 
National Science Foundation, and the SIMBAD database,
operated at CDS, Strasbourg, France. Fig. 1 was created using the
Kapetyn package \citep{KapteynPackage}. We thank the anonymous referee
for their helpful feedback on the original version of the paper.

\bibliographystyle{mnras} \bibliography{phil}

\label{lastpage} \end{document}